\documentstyle[prb,aps]{revtex}

\begin{document}
\draft

\title{
Spin Wave Instability of Itinerant Ferromagnet\\
}

\author{
Takuya {\sc Okabe}
}

\address{
Department of Physics, Kyoto University, Kyoto 606-01
}

\date{
\today
}
\maketitle

\begin{abstract}
We show variationally that instability of the ferromagnetic state
in the Hubbard model is largely controlled by
softening of a long-wavelength spin-wave excitation,
except in the over-doped strong-coupling region where
the individual-particle excitation becomes unstable first.
A similar conclusion is drawn also for the double exchange ferromagnet.
Generally the spin-wave instability may be regarded as a
precursor of the metal-insulator transition.
\end{abstract}


\section{Introduction}
Recently the electron correlation effect
in the strong coupling Hubbard model 
around half filling has been intensively investigated.
Since Nagaoka showed 
the existence of itinerant ferromagnetic ground state
in the limit
$U\rightarrow\infty$ and $n\rightarrow 1$,~\cite{rf:Nagaoka}
several authors~\cite{rf:Roth,rf:RR,rf:Shastry,rf:BE,rf:vdLE,rf:HUM}
attempted variational estimate of 
the stability of the ferromagnetic state in this limit.
On the other side, there are works trying to 
investigate the metal-insulator transition 
from the ferromagnetic side.
In particular, the two-body problem of a particle-hole pair
in the half filled band was treated
as an exactly solvable case of the Mott transition 
by several authors.~\cite{rf:Kohn,rf:DJK1,rf:LJ,rf:vDJ}
To describe physically relevant situations in this approach,
one must treat the many-body problem of the particle-hole bound states.
For example, one may use the BCS-type mean-field approximation
when the ground state is 
magnetically ordered.~\cite{rf:LJ,rf:DJK2,rf:KJSW}

In this paper
we discuss stability of the Nagaoka ferromagnetic state, 
with the problem of the metal-insulator transition in mind. 
We show that the ferromagnetic state in the over-doped region 
of the strong coupling Hubbard model is destabilized by
the individual-particle excitation,
as Shastry et al. noted.~\cite{rf:Shastry}\
However it is found that in almost all the other region
the instability is controlled by softening of the spin-wave stiffness.
In \ref{sec:SFS}, we give a phase diagram showing this feature
on the basis of a variational trial state.
We estimate an upper bound for $\kappa$,
which is defined by $\kappa\equiv zt/U_{\rm c}x$
for the ferromagnetic threshold
in the limit $U\rightarrow\infty$ and $x=1-n\rightarrow 0$.
Stability of the double exchange ferromagnet is discussed
in \ref{sec:Dex}.
Incidentally, in Appendix \ref{sec:n=1},
we discuss that the antiferromagnetic Heisenberg model can be reproduced
by using the results of the two-body problem, or 
from the spin wave in the insulating ferromagnetic state.
Mathematical details
are given in Appendix \ref{ap:Math}.

\section{Variational description of the spin-wave excited state}
\subsection{Random phase approximation}
\label{sec:RPA}
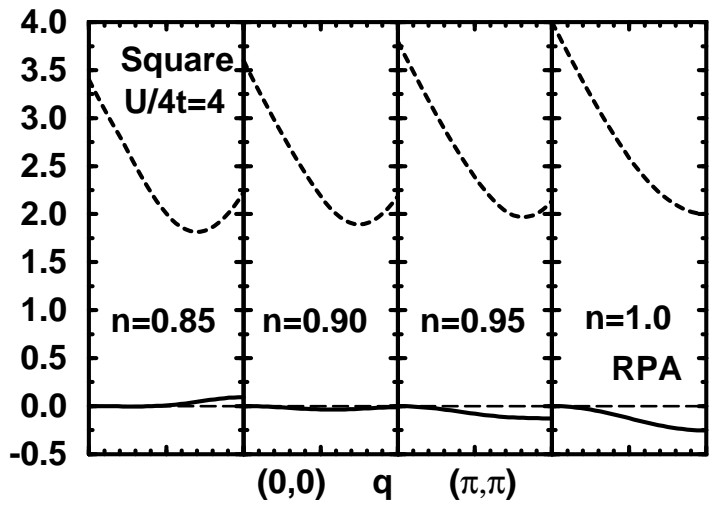
\begin{figure}
\caption{
Dispersion of the spin wave ($\omega_q/4t$)
and the bottom of continuum ($\eta_{q{\rm min}}/4t$)
along $(q,q)$ $(0\le q\le\pi)$ in the RPA.
}
\label{fig:1}
\end{figure}

First investigate the ferromagnetic state,
\begin{equation}
|{\rm F}\rangle\equiv
\prod_{\varepsilon_k<\varepsilon_f} c^\dagger_{k\uparrow}|0\rangle,
\label{RPAF}
\end{equation}
of the Hubbard model,
\begin{eqnarray}
H&=&-\sum_{i,j,\sigma}t_{ij}
c^{\dagger}_{i\sigma}c_{j\sigma}+
U\sum_{i}\hat{n}_{i\uparrow}\hat{n}_{i\downarrow}\nonumber\\
&=&\sum_{k\sigma}\varepsilon_k
c^{\dagger}_{k\sigma}c_{k\sigma}+
U\sum_{i}\hat{n}_{i\uparrow}\hat{n}_{i\downarrow}.
\label{HM}
\end{eqnarray}
By way of illustration, 
here we present results obtained 
in the random phase approximation (RPA) first.
To this end, we may use the following trial state
for the spin-wave excited state;
\begin{equation}
|\Psi_q\rangle=b^\dagger_q|{\rm F}\rangle,
\label{Psiq}
\end{equation}
where
\begin{eqnarray}
b^\dagger_q&\equiv&\frac{1}{\sqrt{L}}
\sum_{i,j}f_q(r_j-r_i)c^\dagger_{i\downarrow}
c_{j\uparrow}e^{iqr_i}
\label{Df:ar}\\
&=&\frac{1}{\sqrt{L}}\sum_k f_q(k)c^\dagger_{k+q\downarrow}c_{k\uparrow},
\label{Df:ak}
\end{eqnarray}
and $L$ denotes the total number of lattice sites.
Then we obtain
\begin{eqnarray}
\langle {\rm F}|b_q[H,b^\dagger_q]|{\rm F}\rangle
&=&\frac{1}{L}\sum_k n_k(\varepsilon_{k+q}-\varepsilon_k+Un)
|f_q(k)|^2\nonumber\\
&&-U\left|\frac{1}{L}\sum_k n_k f_q(k)\right|^2,\label{rpah}\\
\langle {\rm F}|b_qb^\dagger_q|{\rm F}\rangle
&=&\frac{1}{L}\sum_k n_k|f_q(k)|^2,\label{rpan}
\end{eqnarray}
where
\begin{equation}
n_k=\left\{\begin{array}{ll}1,&\qquad\varepsilon_k<\varepsilon_f\\
0,&\qquad\varepsilon_k>\varepsilon_f\end{array}\right.
\end{equation}
and carrier density $n$ is defined by
\[
n\equiv\frac{1}{L}\sum_k\langle {\rm F}| n_k|{\rm F}\rangle.
\]
Taking functional derivative of
\begin{equation}
\omega_q=
\frac{\langle {\rm F}|b_q[H,b^\dagger_q]|{\rm F}\rangle}{
\langle {\rm F}|b_qb^\dagger_q|{\rm F}\rangle},\label{rpaom}
\end{equation}
with respect to $f_q(k)$,
we obtain
\begin{equation}
f_q(k)=\frac{1}{
\varepsilon_{k+q}-\varepsilon_k+Un-\omega_q}
\frac{U}{L}\sum_k n_k f_q(k).
\label{eq:fq}
\end{equation}
Summing $n_k f_q(k)$ over $k$,
we have the eigenequation 
\begin{equation}
\frac{1}{L}\sum_k \frac{n_k}{
\varepsilon_{k+q}-\varepsilon_k+Un-\omega_q}=\frac{1}{U}.
\label{EEforn<1}
\end{equation}
Substituting $f_q(k')=\delta_{kk'}$ in Eqs. (\ref{rpah}),
(\ref{rpan}) and (\ref{rpaom}),
we obtain
\begin{equation}
\eta_q(k)=\varepsilon_{k+q}-\varepsilon_k+Un,
\end{equation}
for the energy of the particle-hole continuum.
The bound state (spin wave) energy $\omega_q$ is given
as a solution of Eq.~(\ref{EEforn<1}).
In particular, for a tight-binding dispersion in a square lattice, 
the results are shown in
Fig.~\ref{fig:1} for ${\bf q}=(q,q)$ $(0\le q\le \pi)$
for various values of $n$.
The spin-wave part of Fig.~\ref{fig:1} is shown in Fig.~\ref{fig:2}.
The result for $n=1$ 
reproduces the two-body result given in Appendix \ref{sec:n=1}.
Discussion based on Eq.~(\ref{EEforn<1})
is equivalent to the random phase approximation
which properly takes into account the two-body correlation effect
of the particle-hole ladder.
In all of the cases shown in the figures,
the ferromagnetic state is unstable to
the spin-wave excitation.

From these results, we observe several points.
(i) $\omega_q$ and $\eta_q(k)$ are separated by energy gap
of order $U$. (Fig.~\ref{fig:1}.)
(ii) The band width of $\omega_q$ becomes narrow
and (iii) the minimum of $\omega_q$ moves away from 
${\bf q}={\bf Q}=(\pi,\pi)$
as density of holes increases (Fig.~\ref{fig:2}).
For example, the minimum of 
$\omega_q$ for $U/4t=4$ and $n=0.9$
lies at ${\bf q}=(\pi,0.3\pi)$.
With respect to (iii),
we are led to the following speculation;
beyond spin-wave instability,
the spin wave with momentum $q=q_{\rm min}$
which gives the minimum $\omega_q\le 0$
will be set to populate the ferromagnetic state,
resulting in Bose-Einstein condensation
of the boson $b^\dagger_{q_{\rm min}}$. 
In particular, around half filling $n\alt 1$,
the resulting state will be 
the commensurate N\'eel ordered phase with ${\bf q}={\bf Q}$.
Upon doping $n\alt 0.93$, e.g. for $U/4t=4$,
the resulting phase will become the spiral state with
incommensurate modulation ${\bf q}\ne {\bf Q}$.
Qualitatively this is consistent with the result of
recent studies.~\cite{rf:spiral,rf:spiral2,rf:spiral3}\
Moreover, further hole doping ($n\alt 0.82$ for $U/4t=4$)
stabilizes the ferromagnetic state,
just as concluded 
from the mean-field treatment.~\cite{rf:spiral2,rf:spiral3}\
However, 
the latter point on the stability of the ferromagnetic state
as well as the above point (i) are shown 
to be modified
by improving the approximation.

\subsection{Improved trial state}
\label{sec:ITS}
Next we consider the following creation operator
to improve the trial state created by Eq.~(\ref{Df:ar});
\begin{equation}
b^\dagger_q\equiv\frac{1}{\sqrt{L}}
\sum_{i,j}f_q(r_j-r_i)
c^\dagger_{i\downarrow}\left(\sin\theta+\cos\theta
c_{i\uparrow}c^\dagger_{i\uparrow}\right)
c_{j\uparrow}e^{iqr_i}.
\label{ImpTrial}
\end{equation}
The wavefunction of this form was first used by
Roth~\cite{rf:Roth} and also adopted later by
Shastry et al.~\cite{rf:Shastry}
to investigate stability of the Nagaoka ferromagnetic state.
However, since the spin wave spectrum for general $q$ and
finite $U$ derived from (\ref{ImpTrial}),
which turns out to be important for our purpose,
has not yet been thoroughly investigated,
we derive results by ourselves from the outset.
Mathematical details are deferred to Appendix \ref{ap:Math}.
Below we show only results
to compare them with those given in the last subsection.

For the square lattice,
the bottom of the continuum $\eta_{q{\rm min}}$
and $\omega_q$
for ${\bf q}=(q,q)$ ($0\le q \le \pi$) and  $U/4t=5$
are shown in Fig.~\ref{fig:3}, 
and $\omega_q$ for various values of $n$
are shown in Fig.~\ref{fig:4}.
These are to be compared with Fig.~\ref{fig:1} and
Fig.~\ref{fig:2}, respectively.
As for $\omega_q$,
in the slightly doped region $1-n\ll 1$,
the results are not modified considerably from those of the RPA.
On the other side, the individual particle-hole
spectrum presents a striking contrast
as is clear from Fig.~\ref{fig:1} and Fig.~\ref{fig:3}.
In the improved estimate, the continuum
no more has energy of order $U$, but
it forms a flat band lying in the low energy region.
This is because of the fact that
vacancy made through the hole doping 
enables particles to hop around,
though it is a quite restricted motion.~\cite{rf:Shastry}
As the density of hole $1-n$ increases,
the band width of $\eta_q$ broadens and
finally we have a vanishing binding energy
$\Delta_q\equiv\eta_{q{\rm min}}-\omega_q$ 
for $q$ which gives the lowest $\omega_q$,
as shown in Fig.~\ref{fig:3}.
In the over-doped region, therefore,
the spin wave for $q\sim q_{\rm min}$
cannot be regarded as a well-defined bound state.

\section{Instability of the ferromagnetic state}
\subsection{Hubbard model}
\label{sec:SFS}
In our previous papers,~\cite{rf:TOdex,rf:TOhund}\
we discussed stability of itinerant ferromagnets
on the basis of a model comprising degenerate orbitals.
In the strong coupling limit, we observed
that the instability condition derived from
the individual particle excitation is more stringent
than the condition concluded for the spin wave instability.
Therefore,
putting more emphasis on the study of
the individual particle excitation
than for the spin wave spectrum,
we calculated the critical interaction $U_{\rm c}$
below which excitation energy of the individual particle-hole pair
becomes negative
as a function of carrier density $n$.~\cite{rf:TOdex}\
The result for the Hubbard model
was that $U_{\rm c}$ approaches a finite value
as $n\rightarrow 1$ both for a square 
and a simple cubic lattice.
In other words,
we could not prove instability of the Nagaoka ferromagnetic state
in the under-doped strong-coupling region,
even though more elaborated trial state than 
that derived from Eq.~(\ref{ImpTrial})
is used to estimate energy of the individual-particle excitation.
On the contrary, for the Hubbard model,
a simple argument can be given, indicating
that $U_{\rm c}$ should become infinity as $n\rightarrow 1$.
To show this,
one may consider the case where holes of concentration $x=1-n$
are doped into the half-filled Hubbard model.
If $x$ is small enough, energy of the complete ferromagnetic state
is given by $-xzt$ per site.
On the other hand,
energy of the antiferromagnetic configuration
is $-2zt^2/U+O(x)$ per site.
Therefore, equating these two,
the critical boundary is expected 
to take the form $zt/U_{\rm c}\propto x$,
as was noted by Nagaoka.~\cite{rf:Nagaoka}\
This argument however does not tell us
whether the instability is brought about locally 
(continuously) or globally.
We show that it is in fact given as
a local instability by investigating the spin wave excitation
for a strong but finite interaction energy.
It is noted here that a quantitative aspect of the above fact
was addressed in a recent work by Hanisch et al,~\cite{rf:HUM}
where emphasis is put on
how far the stable region of the Nagaoka state can be reduced,
rather than its physical origin of our concern.

We display the inverse of the critical coupling
as a function of $n$ in Fig.\ref{fig:5} and Fig.\ref{fig:6}
for a square and a simple cubic lattice,
respectively.~\cite{rf:com2}
In the figures, threshold curves
determined using Eq.~(\ref{ImpTrial}) are shown.
These are calculated by the conditions
$\eta_{\rm min}=0$ (dashed), $\omega_{Q}=0$ (long-dashed)
and $D=0$ (solid line),
where $\eta_{\rm min}$ is the minimum value of $\eta_q(k)$.
In the region above the curves,
the ferromagnetic state is absolutely unstable.
Mathematical details for $\eta_q$, $\omega_q$ and $D$
are given in  Appendix \ref{ap:Math}.
The stiffness constant $D$ for the case $zt/U=0$
was shown as a function of $n$ in Ref.~\cite{rf:TOdex}.
In Fig.\ref{fig:5}, threshold by
$D_{\rm RPA}=0$ using Eq.~(\ref{ap:Drpa})
is also shown.
It is clear how the trial state (\ref{ImpTrial})
improves the result of the RPA;
in the latter we cannot prove instability for any $n$
in the strong coupling region.

The figures show that,
in the strong coupling limit $zt/U=0$,
the individual particle excitation 
(dashed curve) brings about instability
prior to the softening of the stiffness constant
(solid curve),
just in accordance with our previous results.
However the phase boundary in the other region
is primarily determined by the spin-wave instability, $D<0$.
In particular, in the region $U\rightarrow\infty$ and
$n\rightarrow1$,
the phase boundary is of the form $zt/U_{\rm c}=\kappa x$
and is determined by the spin-wave instability.
In this limit, we cannot distinguish the two results by $\omega_Q=0$
and $D=0$,
which in turn do not differ appreciably from $D_{\rm RPA}=0$.
Physically we may say that
the spin-wave instability determines threshold for ferromagnetism
in the region where the metal-insulator transition
is likely to occur.
A similar behavior is observed also for the double exchange ferromagnet
as shown in the next subsection.

As for $\kappa\equiv zt/U_{\rm c}x$,
our calculation gives $\kappa=1.08$ for the square lattice ($z=4$).
For the simple cubic lattice ($z=6$),
we obtained $\kappa=2.08$, which 
is better than $\kappa =3.96$ of Richmond and Rickayzen,~\cite{rf:RR}
who estimated $\kappa$ by assuming a flipped spin 
to stay at a single site,
i.e., not to hop around in a lattice as the spin wave does.
For reference, we cite Nagaoka's estimate~\cite{rf:Nagaoka}
$\kappa=1.47$ for a simple cubic lattice.
This value, however, is not to be compared with our result since
the former does not have a variational significance.
Our result, being based on the variational treatment,
sets the exact upper bound for the
true value of $\kappa$.

\subsection{Double exchange model}
\label{sec:Dex}
Next, we consider
the ferromagnetic Kondo lattice (double exchange) model,
\begin{equation}
H=-t\sum_{i,j,\sigma}c^\dagger_{i\sigma}c_{j\sigma}
+U\sum_i n_{i\uparrow}n_{i\downarrow}
-J'\sum_i {\bf S}_{fi}\cdot{\bf s}_i
+\frac{J'S_f}{2}\sum_{i,\sigma}n_{i\sigma},
\label{fKl}
\end{equation}
where we assume a positive coupling $J'>0$.
This model with $S_f=3/2$ is often used
to describe the lanthanum manganese oxides La$_{1-x}A_x$MnO$_3$,
where $A$ is a divalent ion such as Sr, Pb or Ca.
To create a trial state for the spin wave excitation,
we use the operator 
\begin{equation}
b^\dagger_q\equiv\frac{1}{\sqrt{L}}\sum_{i,j}f_q(r_j-r_i)
\left((c^\dagger_{i\downarrow}c_{j\uparrow}+S_{fi}^-)\sin\theta
+
(c^\dagger_{i\downarrow}c_{i\uparrow}+S_{fi}^-)
c^\dagger_{i\uparrow}c_{j\uparrow}\cos\theta
\right)e^{iqr_i},
\label{ImpTrialdex}
\end{equation}
which reduces to Eq.~(\ref{ImpTrial}) when $S_f=0$.
The expression for $\theta =0$, when $U=J'=\infty$,
was previously treated by us.~\cite{rf:TOdex}

As an example, we show threshold for the
instability of the double exchange ferromagnet,
the ferromagnetic ground state of (\ref{fKl}).
In Fig.\ref{fig:7},
we show $6tn/g_{\rm c}$ determined
by $D=0$ as a function of carrier density $n$ (solid curve)
where $g\equiv JS_f+Un$ represents
the mean-field exchange splitting of the model (\ref{fKl}).
We assumed a tight-binding band in a simple cubic lattice.
In the figure,
we juxtaposed our previous result~\cite{rf:TOdex} (dashed curve),
which was obtained by investigating the instability of
the individual particle excitation; shown as
the dashed curve in the right figure of Fig.~9 of Ref.~\cite{rf:TOdex}.
Experimentally, the itinerant ferromagnetic
state is observed
only in a restricted range $0.2\alt x\equiv 1-n\alt 0.5$.
On the assumption that the model (\ref{fKl})
adequately describes the manganese oxides,
we may conclude that
the observed result is explained by assuming
$\bar{U}/6t\equiv g/6tn \sim 5$,
i.e., $\bar{U}/W\sim 2.5$ with the band width $W=12t$.
In particular,
our results indicate the spin wave instability
in the under-doped regime, while
in the over-doped regime the instability
is controlled by the individual particle excitation
with wavenumber $q\sim k_f$.
Generally the spin wave $\omega_q$
is made unstable first at the momentum
${\bf q}\ne {\bf Q}=(\pi,\pi,\pi)$.
Therefore the resulting phase beyond the instability
is expected to be the incommensurate spiral state,~\cite{rf:IM}
as in the case of the Hubbard model.

In Fig.\ref{fig:8}, we show the $S_f$ dependence of
the phase boundary $4tn/g$ in a square lattice.
Solid curves are determined by $D=0$.
As above,
for the instability due to the individual-particle excitation
(denoted by long-dashed curves),
we adopted the more stringent
condition $\varepsilon_{k=0\downarrow}=\varepsilon_f$
on the basis of our previous results
(7$\cdot$9) and (7$\cdot$10) of Ref.~\cite{rf:TOdex},
than that concluded from the expression derived from (\ref{ImpTrialdex}), 
the counterpart of Eq.~(\ref{Ap:etaqk}).
The figure shows that the threshold near $n\rightarrow 1$
is determined by $D=0$, as in the case of the Hubbard model.
The $D=0$ portion of the phase boundary
increases as a function of $S_f$.
In particular for $S_f=\infty$ in the square lattice,
the boundary is exclusively determined by
the condition $D=0$ down to $n=0$,
while we found that the solid curve $D=0$ and the dashed curve
cross each other around $(n,6tn/g)=(0.5,0.5)$
for $S_f=\infty$ in the simple cubic lattice 
(not shown in Fig.\ref{fig:7}).
Note that in the case $S_f=\infty$ the boundary is the same as
the result of the random phase approximation.
(See $D_{\rm RPA}=0$ of Fig.\ref{fig:5},
as well as Eqs.~(\ref{ap:Drpa}) and ~(\ref{ap:Drpa2}).)
We see that
all the boundaries in Fig.\ref{fig:8} approach to a finite value
as $n\rightarrow 0$.
This is a specific feature of two-dimensional lattices.
Generally for $S_f\ne 0$,
the parmeter $\theta$ in (\ref{ImpTrialdex})
determined variationally
increases from zero to $\pi/2$ as $n$ decreases from 1 to 0.
Thus one can show that 
the condition $D=0$ gives
$4tn/g= 1/\pi$ in the limit $n\rightarrow 0$ for the square lattice,
using Eqs.~(\ref{ap:Drpa}) and ~(\ref{ap:Drpa2}).
On the other hand, in the simple cubic lattice,
we have $1/U\rightarrow 0$ in this limit (Fig.\ref{fig:7}).
Finally we note that the physically relevant situation $S_f=3/2$
lies just in between the classical ($S_f=\infty$)
and the quantum ($S_f=1/2$) limit.

\section{Discussion}
\label{sec:Dis}
In the previous section, we showed that
the ferromagnetic state in the Hubbard model
as well as in the double exchange model
is unstable to the spin-wave excitation in the under-doped region,
while it is unstable to the individual-particle excitation
in the strong-coupling over-doped region.
Generally the spin wave mode $\omega_q$ may take
a mimimum at finite wave vector ${\bf q}={\bf q_0}$.
This minimum is interpreted
to indicate the potential spiral-spin correlation
contained in the ferromagnetic configuration,
which becomes conspicuous as the excitation gap $\omega_{q_0}$ 
approaches zero.
We found that
the softening of the long-wavelength spin wave mode
$D\rightarrow 0$
closely follows the gap collapse $\omega_{q_0}\rightarrow 0$.
Thus we may conclude that the spin-wave instability is
a precursor of the metal-insulator transition.
It is physically plausible that instability of the ferromagnetic state
in the region where the metal-insulator transition is likely to occur
is dictated by the spin wave instability,
since the spin wave is nothing but a particle-hole bound state
in the ferromagnetic vacuum and,
on the other side, we may interpret the Mott insulator as
composed of the particle-hole bound states (Appendix \ref{sec:n=1}).
Therefore we expect that
the above conclusion is generally valid;
for example, instability of ferromagnetism in
the situation appropriate to metallic nickel
will be caused by the individual-particle
excitation since the filling $n\sim 0.2$ per band
is far from being critical for the metal-insulator transition.
In this respect, we note that the Hubbard model
as a model for an itinerant ferromagnet
is a rather exceptional case,
because 
the ferromagnetic state if any
can be realized only around half-filling (Nagaoka limit).
This is the reason why we found a dominant role
played by the spin-wave instability in
Fig.\ref{fig:5} and Fig.\ref{fig:6}.

Finally let us speculate on 
the approach to the metal-insulator transition
from the ferromagnetic side.
We could derive
the antiferromagnetic Heisenberg model
from the spin-wave dispersion of the
ferromagnetic insulating state,
and observed that, unlike the individual-particle excitation spectrum,
the spin-wave dispersion itself does not
change drastically upon hole doping
(Fig.~\ref{fig:1} and  Fig.~\ref{fig:3}).
Then 
if the spin wave as a particle-hole bound state
is robust even when slight amount of holes are doped,
the paramagnetic phase realized after the spin-wave instability 
may show anomalous metallic properties.
This kind of consideration make sense 
just around half filling only where
the lowest energy spin-wave state can be regarded
as a well-defined bound state (Fig.~\ref{fig:3}).
Although it is an interesting problem
to consider the interacting 
ferromagnetic spin wave as an elementary constituent,
further investigation on this point requires
complicated calculation which is far beyond the scope of this article.

\section*{Acknowledgments}
The author would like to thank Professor K. Yamada for
discussions and critical comments.
This work is supported  by
Research Fellowships of the Japan Society for the
Promotion of Science for Young Scientists.

\appendix

\section{Two-body problem and the Heisenberg model}
\label{sec:n=1}
Let us consider the simplest case of a particle-hole pair in
the ferromagnetic band insulator,
\begin{equation}
|{\rm F}\rangle\equiv\prod_k c^\dagger_{k\uparrow}|0\rangle.
\label{TBPF}
\end{equation}
Then the eigenequation for $\omega_q$,
which is the eigenvalue of the state (\ref{Psiq}), is given by
\begin{equation}
\frac{1}{L}\sum_k \frac{1}{
\varepsilon_{k+q}-\varepsilon_k+U-\omega_q}=\frac{1}{U}.
\label{2EE}
\end{equation}
This is obtained as a limit $n\rightarrow 1$ of Eq.~(\ref{EEforn<1}),
and is the exact result
of the two-body problem.~\cite{rf:Kohn,rf:DJK1,rf:LJ,rf:vDJ}
It is also derived as the condition for the particle-hole ladder
to have a pole.
We note that 
the bound state solution $\omega_q$ is nothing but a spin wave
in the ferromagnetic vacuum (\ref{TBPF}).
On the other side, the internal structure of the bound state
$f_q(k)$ is given by
\begin{equation}
f_q(k)= \frac{1}{\sqrt{L}}\cdot\frac{U}{
\varepsilon_{k+q}-\varepsilon_k+U-\omega_q},
\label{eq:fq2}
\end{equation}
without loss of generality.

We investigate a general case of the tight-binding band,
for which $\varepsilon_k$ is given by
\begin{equation}
\varepsilon_k=-t\sum_{\bar{\delta}} e^{ik\bar{\delta}},
\label{TBdisp}
\end{equation}
where the sum is taken over nearest-neighbor vectors $\bar{\delta}$.
To the accuracy of order $O(t/U)$,
from Eq.~(\ref{2EE}) we obtain an expression for $\omega_q$,
\begin{eqnarray}
\omega_q&=&
-\frac{1}{U}\frac{1}{L}\sum_k (\varepsilon_{k+q}-\varepsilon_k)^2\\
&=&-\frac{2t^2}{U}
\sum_{\bar{\delta}}
\left(1-e^{iq\bar{\delta}}\right),
\label{omq}
\end{eqnarray}
where Eq.~(\ref{TBdisp}) is substituted.
The case $n=1$ of Fig.~\ref{fig:2} and Fig.~\ref{fig:4}
can be well fitted by this expression.

As for $f_q(k)$, we have
\begin{equation}
f_q(k)= \frac{1}{\sqrt{L}}
\left(1-\frac{\varepsilon_{k+q}-\varepsilon_k}{U}\right),
\end{equation}
to the accuracy of order $O(t/U)$.
We then obtain
\begin{eqnarray}
f_q(r_j-r_i)&=&\frac{1}{\sqrt{L}}\sum_k f_q(k)e^{-ik(r_j-r_i)}\nonumber\\
&=&\delta_{r_j-r_i}+\frac{t}{U}\sum_{\bar{\delta}}
 \delta_{r_j-r_i-\bar{\delta}}
\left(e^{iq\bar{\delta}}-1\right),
\label{eq:fq3}
\end{eqnarray}
and
\begin{eqnarray}
b^\dagger_i&=&\frac{1}{\sqrt{L}}\sum_q b^\dagger_qe^{-iqr_i}\nonumber\\
&=&\frac{1}{L}\sum_q \sum_{i',j'}
f_q(r_{j'}-r_{i'})
c^\dagger_{i'\downarrow}c_{j'\uparrow}e^{iq(r_{i'}-r_i)}\nonumber\\
&=&c^\dagger_{i\downarrow}c_{i\uparrow}
+\frac{t}{U}\sum_{\bar{\delta}}
\left(c^\dagger_{i-\bar{\delta}\downarrow}c_{i\uparrow}
-c^\dagger_{i\downarrow}c_{i+\bar{\delta}\uparrow}
\right).
\label{ai}
\end{eqnarray}
The result Eq.~(\ref{ai}) shows that
the boson $b^\dagger_i$
for non-zero $t/U$ has an internal structure 
extending to neighboring sites of the site where it is created.
Physically this structure is interpreted as a singlet cloud
formed with the neighboring sites.

Energy of the localized boson is given by
\begin{equation}
\omega_i=
\frac{\langle {\rm F}|b_i[H,b^\dagger_i]|{\rm F}\rangle}
{\langle {\rm F}|b_ib^\dagger_i|{\rm F}\rangle}
=-\frac{2zt^2}{U},
\label{omi}
\end{equation}
where $z$ is a coordination number, \(
z=\sum_{\bar{\delta}}1.
\)
The result Eq.~(\ref{omi}) is obtained also as the center of gravity
of the band $\omega_q$, Eq.~(\ref{omq}),
\begin{equation}
\bar{\omega}_q=\frac{1}{L}\sum_q \omega_q=
-\frac{2zt^2}{U}.
\end{equation}
From the form of the structure, Eq.~(\ref{ai}),
Eq.~(\ref{omi}) is interpreted as $-z\cdot 2t^2/U$,
i.e., as a sum of energy of singlets formed with
the $z$ nearest neighbors.
The factor $2$ is due to the two processes due to
a particle and a hole hopping, which are represented in
the two terms in the parenthesis of Eq.~(\ref{ai}).

Now let us introduce the creation and annihilation operator of
the hard-core boson $\tilde{b}^\dagger_i$ and $\tilde{b}_i$
in place of $b^\dagger_i$ and $b_i$;
\begin{equation}
b^\dagger_i\rightarrow\tilde{b}^\dagger_i,\qquad
b_i\rightarrow\tilde{b}_i.
\label{a>tila}
\end{equation}
The bound state created by $b^\dagger_i$
extends only to the nearest-neighboring sites of $i$,
as indicated from Eq.~(\ref{ai}).
As a result,
total energy of two localized bosons differs from $2\omega_i$
only when they are in the nearest-neighboring sites,
when the total energy amounts to $-4(z-1)t^2/U$.
Increase by an amount $2J\equiv 4t^2/U$ from $2\omega_i$
is due to overlap of the singlet cloud of the two bosons.
In the hard-core boson picture
this must be regarded as an interaction energy,
i.e., the interaction part of the Hamiltonian for the hard-core boson
is given by
\begin{equation}
\tilde{V}=2J\sum_{\langle i,j\rangle}
\tilde{b}^\dagger_i\tilde{b}_i
\tilde{b}^\dagger_{j}\tilde{b}_{j},
\label{Vhcb}
\end{equation}
where the sum is taken over the nearest-neighbor pairs.
On the other side for the hopping part,
the one-body energy $\omega_q$ is given by Eq.~(\ref{omq}).
In terms of $J\equiv 2t^2/U$ 
we rewrite it as
\begin{eqnarray}
\tilde{T}&=&\sum_q \omega_q \tilde{b}^\dagger_q\tilde{b}_q\nonumber\\
&=&-J\sum_q \sum_{\bar{\delta}}\left(1-e^{iq\bar{\delta}}\right)
\tilde{b}^\dagger_q\tilde{b}_q\nonumber\\
&=&J\sum_{\langle i,j\rangle}
\left(-
\tilde{b}^\dagger_i\tilde{b}_i-\tilde{b}^\dagger_j\tilde{b}_j
+
\tilde{b}_i\tilde{b}^\dagger_j
+\tilde{b}_i^\dagger \tilde{b}_j\right).
\label{Thcb}
\end{eqnarray}

Instead of the hard-core boson,
we can equivalently use the quantum operator for the spin $S=1/2$,
which are defined by
\begin{eqnarray}
S_{zi}&=&\frac{1}{2}-\tilde{b}^\dagger_i\tilde{b}_i,\nonumber\\
S^+_{i}&=&\tilde{b}_i,\qquad
S^-_{i}=\tilde{b}^\dagger_i.
\label{S2tila}
\end{eqnarray}
Then as the effective model to describe
the half-filled Hubbard model in the strong coupling regime,
we can reproduce the antiferromagnetic Heisenberg model
in terms of these spin operators:
Putting Eq.~(\ref{Vhcb}) and Eq.~(\ref{Thcb}) together,
we obtain
\begin{eqnarray}
\tilde{H}&=&\tilde{T}+\tilde{V}\nonumber\\
&=&2J\sum_{\langle i,j\rangle}
\left(\bigl(\frac{1}{2}-\tilde{b}^\dagger_i\tilde{b}_i\bigr)
\bigl(\frac{1}{2}-\tilde{b}^\dagger_j\tilde{b}_j\bigr)
+\frac{1}{2}\left(
\tilde{b}_i\tilde{b}^\dagger_j+
\tilde{b}_i^\dagger \tilde{b}_j\right)-\frac{1}{4}\right)\nonumber\\
&=&2J\sum_{\langle i,j\rangle}
\left(S_{zi}S_{zj}+\frac{1}{2}\left(
S^+_{i}S^-_{j}+S^-_{i}S^+_{j}\right)-\frac{1}{4}\right)\nonumber\\
&=&2J\sum_{\langle i,j\rangle}
\left({\bf S}_i \cdot {\bf S}_{j}-\frac{1}{4}\right).
\label{AFHM}
\end{eqnarray}

Similarly as above,
as an effective model for the case when holes are doped,
the following Hamiltonian is suggested; 
\begin{equation}
H=t\sum_{i,j}\tilde{f}^\dagger_{i}\tilde{f}_{j}
-t\sum_{i,j}
\tilde{b}^\dagger_{i}\tilde{f}_{i}\tilde{f}^\dagger_{j}\tilde{b}_{j}
+2J\sum_{\langle i,j\rangle}
\left({\bf S}_i \cdot {\bf S}_{j}-\frac{1}{4}\right).
\label{tJ}
\end{equation}
The first term describes the hopping process of a doped hole,
$\tilde{f}^\dagger_{i}\equiv c_{i\uparrow}(
1-\tilde{b}^\dagger_i\tilde{b}_i)$, and
the second term takes into account the hopping of the boson 
when its neighboring sites are vacant.
The Hilbert space of Eq.~(\ref{tJ})
is spanned at each site
by $|\uparrow_i\rangle$,
$|\downarrow_i\rangle\equiv\tilde{b}^\dagger_i|\uparrow_i\rangle$
and vacancy $|0_i\rangle\equiv \tilde{c}_{i\uparrow}|\uparrow_i\rangle$.
The fermion operator $\tilde{c}_{i\uparrow}$ must thus be operated
on the sites where there is no boson.
To take this into account, one may use $\tilde{f}^\dagger_{i}$
instead of  $\tilde{c}_{i\uparrow}$.

\section{Calculation for 
the improved variational state}
\label{ap:Math}
We calculate $\omega_q$ 
for the creation operator,
\begin{eqnarray}
b^\dagger_q&\equiv&\frac{1}{\sqrt{L}}
\sum_{i,j}\left(f_q(r_j-r_i)c^\dagger_{i\downarrow}c_{j\uparrow}
+\bar{f}_q(r_j-r_i)c^\dagger_{i\downarrow}c_{i\uparrow}
c^\dagger_{i\uparrow}c_{j\uparrow}
\right)e^{iqr_i}.
\label{Ap:ImpTrial}
\end{eqnarray}
To obtain results, 
we may follow the same procedure as given in \ref{sec:RPA}.
First we obtain
\begin{eqnarray}
\langle {\rm F}|b_q[H,b^\dagger_q]|{\rm F}\rangle
&=&\frac{1}{L}\sum_k n_k(\varepsilon_{k+q}-\varepsilon_k+Un)
|f_k|^2
-U\left|\frac{1}{L}\sum_k n_k f_k\right|^2
%
%
+\frac{1}{L}\sum_k \Delta_{kq}|\bar{f}_k|^2
+\frac{1}{L^2}\sum_{k,p}\Gamma_{kpq}\bar{f}^*_k \bar{f}_p
\nonumber\\
&&+\left(\frac{1}{L}\sum_k n_k
(\varepsilon_{k+q}-\varepsilon_k)\bar{f}^*_k
\right)
\left(\frac{1}{L}\sum_k n_k f_k\right)
+\left[\begin{array}{c}
\bar{f}^*_k\rightarrow f^*_k\\
f_k\rightarrow \bar{f}_k
\end{array}\right]
\nonumber\\
&&+\frac{1-n}{L}\sum_k n_k (\varepsilon_{k+q}-\varepsilon_k)
\bar{f}^*_k f_k
+\left[\begin{array}{c}
\bar{f}^*_k\rightarrow f^*_k\\
f_k\rightarrow \bar{f}_k
\end{array}\right],
\label{Ap:Ham}
\end{eqnarray}
and
\begin{eqnarray}
\langle {\rm F}|b_qb^\dagger_q|{\rm F}\rangle
&=&\frac{1}{L}\sum_k n_k|f_k|^2
+\frac{1-n}{L}\sum_k |\bar{f}_k|^2
+\left|\frac{1}{L}\sum_{k}n_k  \bar{f}_k\right|^2
%
%
+\left(\frac{1}{L}\sum_k n_k\bar{f}^*_k\right)
\left(\frac{1}{L}\sum_k n_kf_k\right)
+\left[\begin{array}{c}
\bar{f}^*_k\rightarrow f^*_k\\
f_k\rightarrow \bar{f}_k
\end{array}\right]
\nonumber\\
&&+\frac{1-n}{L}\sum_k n_k \bar{f}^*_k f_k
+\left[\begin{array}{c}
\bar{f}^*_k\rightarrow f^*_k\\
f_k\rightarrow \bar{f}_k
\end{array}\right].
\label{Ap:Norm}
\end{eqnarray}
Here we denoted $f_q(k)$ simply as $f_k$.
Bracketed expressions mean to repeat the preceding
terms with the replaced functions as indicated in the brackets.
$\Delta_{kq}$ and $\Gamma_{kpq}$
in Eq.~(\ref{Ap:Ham}) is
the case $S_f=0$ of the expression 
defined in (4$\cdot$6) and (4$\cdot$7) of Ref.~\cite{rf:TOdex}.
In the tight-binding dispersion Eq.~(\ref{TBdisp}),
the former is given by
\begin{equation}
\Delta_{kq}=|\epsilon_g|-(1-n)\varepsilon_k
+\left((1-n)^2-\left|\frac{\epsilon_g}{zt}\right|^2\right)\varepsilon_{k+q},
\label{Ap:delkq}
\end{equation}
where
\begin{equation}
\epsilon_g\equiv\frac{1}{L}\sum_k n_k \varepsilon_k.
\end{equation}

For the wavefunction Eq.~(\ref{ImpTrial}),
one may replace $f_k$ and $\bar{f}_k$
in the above expressions by
$f_k\sin\theta$ and $f_k\cos\theta$, respectively.
A function $f_k$ and a parameter $\theta$ have to be
fixed so as to minimize $\omega_q$.
We note that Eq.~(\ref{ImpTrial}) for $\theta=\pi/2$
gives Eq.~(\ref{Df:ar}).
Moreover, Eq.~(\ref{ImpTrial}) 
becomes the trial state for 
the case $U=\infty$~\cite{rf:Roth,rf:Shastry}\
by assuming $\theta=0$,
since in this case
the variational state does not depend on the interaction energy $U$ at all.
For a finite value of $U$, 
the parameter $\theta$
takes a value in the range $0\le \theta\le\pi/2$.

\subsection{Individual particle excitation}
To derive excitation energy
of the individual particle-hole pair,
we may set $f_{k'}=\delta_{kk'}$ in
Eqs.~(\ref{Ap:Ham}) and (\ref{Ap:Norm})
to calculate \[
\omega_q=
\frac{\langle {\rm F}|b_q[H,b^\dagger_q]|{\rm F}\rangle}{
\langle {\rm F}|b_qb^\dagger_q|{\rm F}\rangle.}\]
Then we obtain
\begin{equation}
\eta_q(k,\theta)=
\frac{\Delta_{kq}\cos^2\theta
+(\varepsilon_{k+q}-\varepsilon_k+Un)\sin^2\theta
+2(1-n)(\varepsilon_{k+q}-\varepsilon_k)\sin\theta\cos\theta}
{(1-n)\cos^2\theta+2(1-n)\sin\theta\cos\theta+\sin^2\theta}.
\label{Ap:etaqk}
\end{equation}
For $\eta_q(k)$
this expression should be minimized with respect to $\theta$,
and $\eta_{\rm min}$ is defined as a minimum of $\eta_q(k)$.

\subsection{Spin wave dispersion}
To minimize $\omega_q$ with respect to $f_k$,
we take functional derivative $\partial\omega_q/\partial f^*_p$
after replacing $f_k$ and $\bar{f}_k$ in 
Eqs.~(\ref{Ap:Ham}) and ~(\ref{Ap:Norm})
by $f_k\sin\theta$ and $f_k\cos\theta$.
Then 
we obtain an equation,
\begin{eqnarray}
&&\left(\Delta_{kq} f_k+
\frac{1}{L}\sum_p \Gamma_{kpq}f_p\right)\cos^2\theta
+\left((\varepsilon_{k+q}-\varepsilon_k+Un)f_k
-\frac{U}{L}\sum_k n_k f_k\right) \sin^2\theta+\nonumber\\
&&\qquad+\biggl(\frac{\varepsilon_{k+q}-\varepsilon_k}{L}\sum_k n_k f_k
+\frac{1}{L}\sum_k n_k (\varepsilon_{k+q}-\varepsilon_k)f_k
%
%
+2(1-n)(\varepsilon_{k+q}-\varepsilon_k)f_k
\biggr)\sin\theta\cos\theta
\nonumber\\
&=&\omega_q\biggl[\left(
(1-n)f_k+\frac{1}{L}\sum_k n_k f_k\right)\cos^2\theta
+f_k \sin^2\theta
%
%
+\left(\frac{2}{L}\sum_k n_k f_k +2(1-n)f_k\right)\sin\theta\cos\theta
\biggr].
\label{Ap:Eigeneq}
\end{eqnarray}

We investigate the tight-binding model in
a square ($d=2$) and simple cubic lattice ($d=3$),
for which
\begin{equation}
\varepsilon_k=-\frac{1}{d}\sum_{i=1}^d \cos(k_i).
\end{equation}
Here and below we set $zt=1$ where $z=2d$.
Furthermore, we are interested in the dispersion $\omega_q$
for $q$ along the diagonal of the Brillouin zone,
i.e., for ${\bf q}=(q,q)$
and ${\bf q}=(q,q,q)$  ($0\le q\le\pi$) in the square
and the simple cubic lattice, respectively.
Then we can cast Eq.~(\ref{Ap:Eigeneq}) into the following form,
\begin{equation}
D(k,q)f_q(k)=\sum_{i=1}^3N_i(k,q)F_i(q),
\label{Ap:f}
\end{equation}
where
\begin{equation}
D(k,q)=\left(\Delta_{kq}-(1-n)\omega_q\right)\cos^2\theta
+\left(\varepsilon_{k+q}-\varepsilon_k+Un-\omega_q\right)\sin^2\theta
+2(1-n)(\varepsilon_{k+q}-\varepsilon_k-\omega_q)
\sin\theta\cos\theta,
\end{equation}
\begin{equation}
F_i(q)\equiv\frac{1}{L}\sum_k n_k(\delta_{i1}-\varepsilon_k
\delta_{i2}+\tilde{\varepsilon}_k\delta_{i3}
)f_q(k),\label{Ap:F}
\end{equation}
\begin{equation}
N_1(k,q)\equiv\left(
\omega_q-|\epsilon_g|\varepsilon_q-(1-n)\varepsilon_{k+q}\right)
\cos^2\theta+U\sin^2\theta
+(2\omega_q-\varepsilon_{k+q}+\varepsilon_k)\sin\theta\cos\theta,
\end{equation}
\begin{equation}
N_2(k,q)\equiv-\left(
|\epsilon_g|\varepsilon_{k+q}+(1-n)\varepsilon_q\right)\cos^2\theta
-(1+\varepsilon_q)\sin\theta\cos\theta,
\end{equation}
\begin{equation}
N_3(k,q)\equiv-\left(
|\epsilon_g|\tilde{\varepsilon}_{k+q}
+(1-n)\tilde{\varepsilon}_q\right)\cos^2\theta
-v_q\sin\theta\cos\theta.
\end{equation}
Here we introduced 
\begin{equation}
\tilde{\varepsilon}_k=\frac{1}{d}\sum_{i=1}^d \sin(k_i),
\qquad
v_q=\sin(q).
\end{equation}

Solving Eq.~(\ref{Ap:f}) for $f_q(k)$ and
substituting the result into Eq.~(\ref{Ap:F}),
we get the eigenequation
\begin{equation}
\det\left(A_{ij}(q)-\delta_{ij}\right)=0,
\label{Ap:det=0}
\end{equation}
where
\begin{equation}
A_{ij}(q)=\frac{1}{L}\sum_k n_k(\delta_{i1}-\varepsilon_k\delta_{i2}
+\tilde{\varepsilon}_k\delta_{i3})
\frac{N_j(k,q)}{D(k,q)}.
\end{equation}
To obtain the spin wave energy $\omega_q$,
we must minimize the solution 
$\omega_q(\theta)$ of Eq.~(\ref{Ap:det=0})
with respect to $\theta$.

\subsection{Spin wave stiffness constant}
\label{Ap:SwsofKl}
The spin-wave stiffness constant $D$ is defined by
\begin{equation}
\omega_q=Dq^2.\qquad\qquad(q\rightarrow 0)
\end{equation}
In the long wavelength limit $q\rightarrow 0$,
we can expand $f_q(k)$ with respect to $q$
and derive $D$ analytically from 
Eq.~(\ref{Ap:Ham}) and Eq.~(\ref{Ap:Norm}).
Below we give only the resulting expressions.

As shown in our previous paper,~\cite{rf:TOdex}\
the results for the Hubbard model are obtained
as a special case of those for the Kondo lattice model (\ref{fKl}).
Using Eq.~(\ref{ImpTrialdex}),
we calculated $D$ for this general model;
\begin{equation}
D(\theta)=D_0-\delta D(\theta),
\end{equation}
where \(
D_0=|\epsilon_g|/2Sz
\) in terms of twice the spontaneous magnetization $2S\equiv 2S_f+n$, 
and
\begin{equation}
\delta D(\theta)=
\frac{I
\left(
(1-n)\cos^2\theta+\sin^2\theta+(2-n)\sin\theta\cos\theta\right)^2
}{2S(1+|\epsilon_g|zI\cos^2\theta /2)(1+2\sin\theta\cos\theta
)}
\qquad (>0).
\end{equation}
In these expressions, $I$, $v_{k_x}$ and $\Delta_{k0}$
are given by
\begin{equation}
I=\frac{1}{L}\sum_k  \frac{n_kv^2_{k_x}}
{\Delta_{k0}\cos^2\theta+g\sin^2\theta},
\label{Ap:I}
\end{equation}
\begin{equation}
v_{k_x}=\frac{\partial \varepsilon_k}{\partial k_x},
\end{equation}
and
\begin{equation}
\Delta_{k0}=(2S_f+1)\left(|\epsilon_g|-(1-n)\varepsilon_k\right)
+\left((1-n)^2-|\epsilon_g|^2\right)\varepsilon_{k}
\qquad (>0).
\end{equation}
A parameter $g$ in Eq.~(\ref{Ap:I}) is defined by
$g=Un+J'S_f$, and represents
the mean-field exchange splitting of the model (\ref{fKl}).
We must minimize $D(\theta)$ to obtain $D$.
To reproduce the result for the Hubbard model,
we may set $S_f=0$.
The particular case of the above results,
i.e., in the strong coupling limit $g=\infty$ ($\theta=0$),
was obtained previously.~\cite{rf:Shastry,rf:TOdex}\
For the Hubbard model,
the result of the random phase approximation $D_{\rm RPA}$ is
obtained by setting $S_f=0$ and $\theta=\pi/2$, 
\begin{equation}
D_{\rm RPA}=
\frac{|\epsilon_g|}{zn}-\frac{1}{Un^2}
\frac{1}{L}\sum_k  n_kv^2_{k_x}.
\label{ap:Drpa}
\end{equation}
A similar expression is obtained in the limit $S\rightarrow \infty$
where we should assume $\theta=\pi/2$
to keep $I$ and $2S \delta D(\theta)$ finite.
Then we obtain
\begin{equation}
D_{S\rightarrow\infty}=\frac{1}{2S}\left(
\frac{|\epsilon_g|}{z}-\frac{1}{g}
\frac{1}{L}\sum_k  n_kv^2_{k_x}\right).
\label{ap:Drpa2}
\end{equation}
In the main text,
we regarded the approximation scheme giving the result
corresponding to (\ref{ap:Drpa}) and (\ref{ap:Drpa2})
as the random phase approximation.
The former result (\ref{ap:Drpa}) is
well known for the Hubbard model.
The latter was recently obtained by the other method.~\cite{rf:Fu}

%


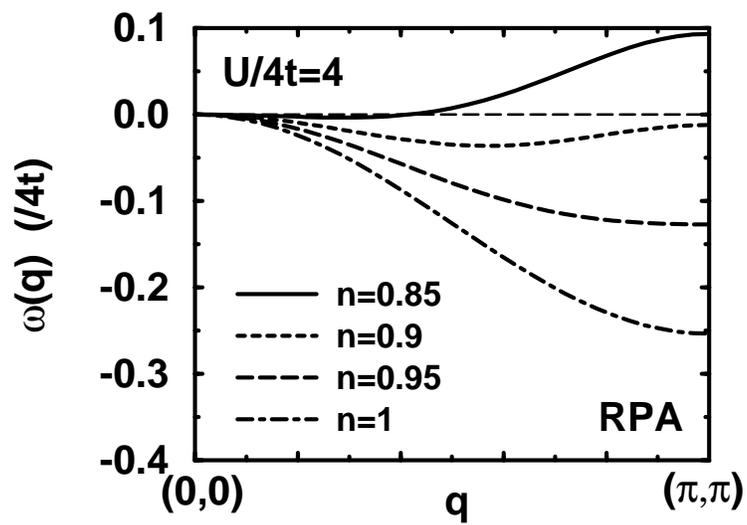
\begin{figure}
\caption{
Dispersion $\omega(q)$
for $U/4t=4$ in the RPA.
}
\label{fig:2}
\end{figure}

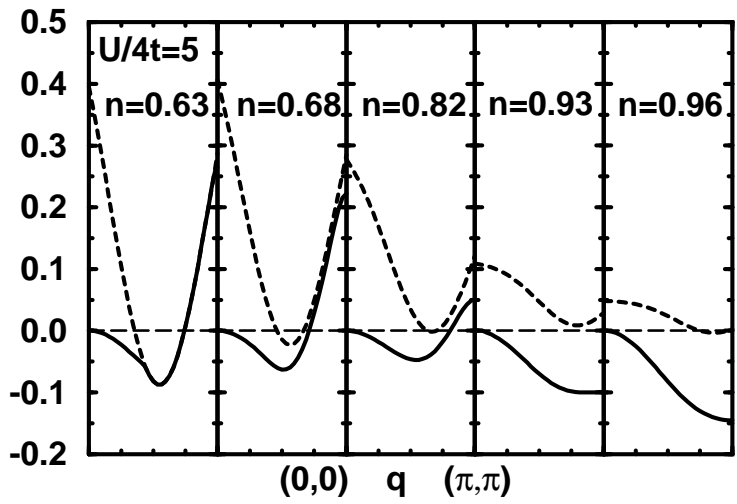
\begin{figure}
\caption{
Dispersion of the spin wave ($\omega_q/4t$)
and the bottom of continuum ($\eta_{q{\rm min}}/4t$)
along $(q,q)$ $(0\le q\le\pi)$ calculated
with the improved trial state.
}
\label{fig:3}
\end{figure}
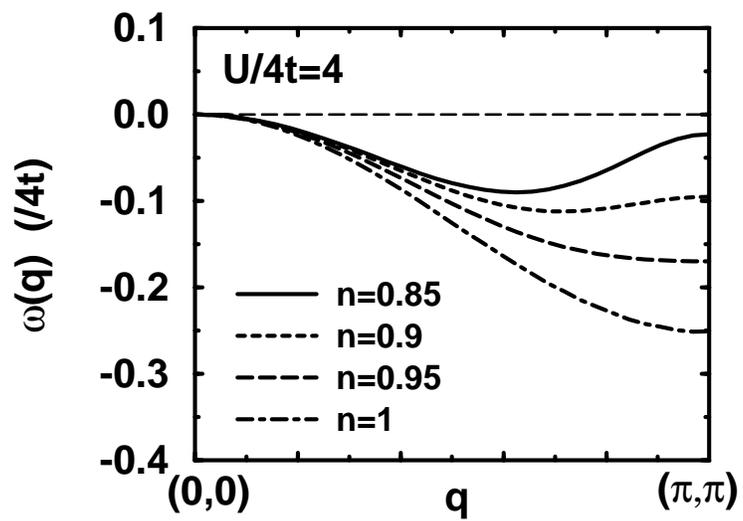
\begin{figure}
\caption{
Dispersion $\omega(q)$
for $U/4t=4$ for the improved trial state.
}
\label{fig:4}
\end{figure}

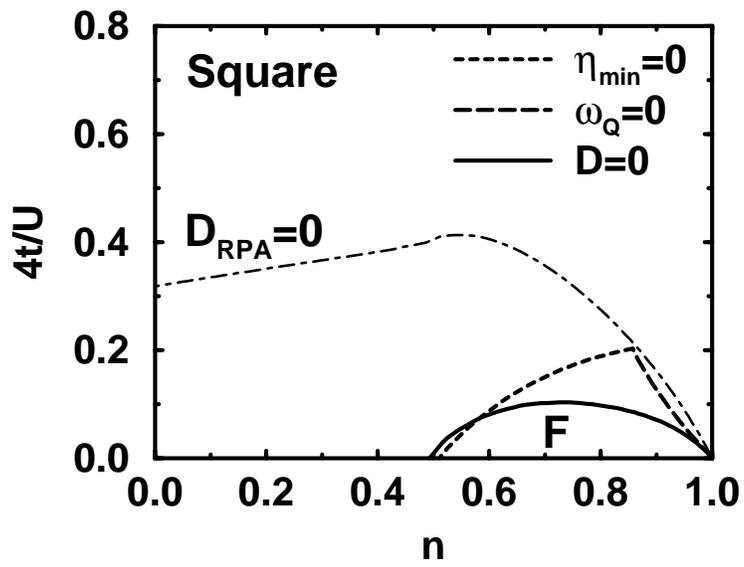
\begin{figure}
\caption{
Threshold for the stability of
the ferromagnetic state in a square lattice.
}
\label{fig:5}
\end{figure}
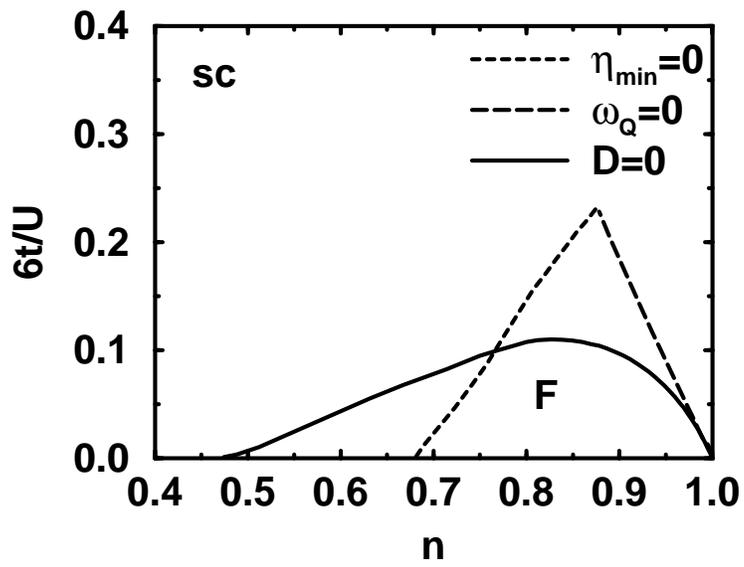
\begin{figure}
\caption{
Threshold for the stability of
the ferromagnetic state in a simple cubic lattice.
}
\label{fig:6}
\end{figure}
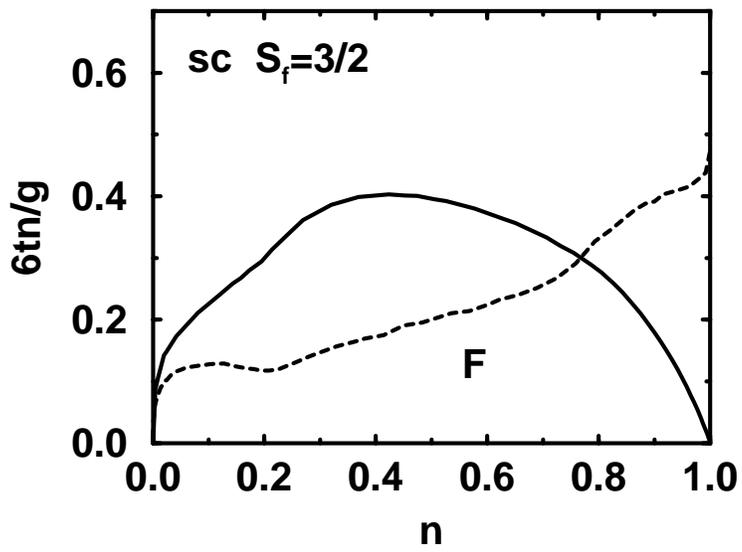
\begin{figure}
\caption{Inverse of the critical coupling
 $6tn/g_{\rm c}$ as a function of $n$ for the 
$S_f=3/2$ ferromagnetic Kondo lattice model in a simple cubic lattice.
The ferromagnetic state is unstable outside the region denoted by
`F'.
}
\label{fig:7}
\end{figure}

\begin{figure}
\caption{
 $4tn/g_{\rm c}$ as a function of $n$ for the 
ferromagnetic Kondo lattice model in a square lattice
for $S_f=1/2$, 3/2 and $\infty$.
Solid curves are determined by $D=0$.
Instability of the individual-particle excitation occurs 
above the long-dashed curves.
The ferromagnetic state is unstable outside the region denoted by
`Ferromagnetic'.
}
\label{fig:8}
\end{figure}


\begin{references}



\bibitem{rf:Nagaoka}Y. Nagaoka, Phys. Rev. {\bf 147}, 392 (1966).
\bibitem{rf:Roth}L. M. Roth,
J. Phys. Chem. Solids {\bf 28}, 1549 (1967);
J. Appl. Phys. {\bf 39}, 474 (1968).
\bibitem{rf:RR}P. Richmond and G. Rickayzen,
J. Phys. {\bf C2}, 528 (1969).
\bibitem{rf:Shastry}B. S. Shastry, H. R. Krishnamurthy and P. W. Anderson,
Phys. Rev. {\bf B41}, 2375 (1990).

\bibitem{rf:BE}A. G. Basile and V. Esler, Phys. Rev. {\bf B41},
4842 (1990).
\bibitem{rf:vdLE}W. von der Linden and D. M. Edwards,
J. Phys.: Condens. Matter {\bf 3}, 4917 (1991).

\bibitem{rf:HUM}T. Hanisch, G. S. Uhrig and 
E. M\"uller-Hartmann, cond-mat 9707286.


\bibitem{rf:Kohn}W. Kohn, Phys. Rev. {\bf 133}, A171 (1964).
\bibitem{rf:DJK1}K. Dichtel, R. J. Jelitto and H. Koppe,
Z. Phys. {\bf B246}, 248 (1971).
\bibitem{rf:LJ}Q. P. Li and R. Joynt, Phys. Rev. {\bf B47}, 3979 (1993);
{\bf B49}, 1632 (1994).
\bibitem{rf:vDJ}P. G. J. van Dongen and V. Jani\v s,
Phys. Rev. Lett. {\bf 72}, 3258 (1994).


\bibitem{rf:DJK2}
K. Dichtel, R. J. Jelitto and H. Koppe,
Z. Phys. {\bf B251}, 173 (1972).
\bibitem{rf:KJSW}H. R. Krishnamurthy, C. Jayaprakash,
S. Sarker and W. Wenzel, Phys. Rev. Lett. {\bf 64}, 950 (1990);
S. Sarker, C. Jayaprakash, H. R. Krishnamurthy and W. Wenzel,
Phys. Rev. {\bf B43}, 8775 (1991);
C. Jayaprakash, H. R. Krishnamurthy, S. Sarker and W. Wenzel,
Europhys. Lett. {\bf 15}, 625 (1991).

\bibitem{rf:spiral}B. I. Shraiman and E. D. Siggia,
Phys. Rev. Lett. {\bf 62}, 1564 (1989).
\bibitem{rf:spiral2}C. Jayaprakash, H. R. Krishnamurthy and S. Sarker,
Phys. Rev. {\bf B40}, 2610 (1989).
\bibitem{rf:spiral3}D. Yoshioka,
J. Phys. Soc. Jpn. {\bf 58}, 1516 (1989).

\bibitem{rf:TOdex}T. Okabe, Prog. Theor. Phys. {\bf 97}, 21 (1997);
{\bf 97}, 559 (1997).
\bibitem{rf:TOhund}T. Okabe, Prog. Theor. Phys. {\bf 98}, No.2 (1997).

\bibitem{rf:com2}The same kind of phase diagram was given
in Ref.~\cite{rf:Shastry},
where the result of Ref.~\cite{rf:RR} had to be cited
to supplement the portion $zt/U_{\rm c}\propto x$ for
$U\rightarrow\infty$ and $x\rightarrow 0$.


\bibitem{rf:IM}
J. Inoue and S. Maekawa, Phys. Rev. Lett. {\bf 74}, 3407 (1995).
\bibitem{rf:Fu}
N. Furukawa, J. Phys. Soc. Jpn. {\bf 65}, 1174 (1996). 

\end{references}
\end{document}